\documentclass[a4paper]{jpconf}
\usepackage{graphicx}
\begin{document}
\title{Wide binaries as a critical test for Gravity theories}

\author{X Hernandez, M A. Jim\'enez and C Allen}

\address{Instituto de Astronom\'{\i}a, Universidad Nacional Aut\'onoma de M\'exico, AP 70-264, Distrito Federal 04510, M\'exico}

\ead{xavier@astro.una,.mx}

\begin{abstract}
Assuming Newton's gravity and GR to be valid at all scales leads to the dark matter hypothesis as a 
requirement demanded by the observed dynamics and measured baryonic content at galactic and extragalactic scales.
Alternatively, modified gravity scenarios where a change of regime appears at acceleration scales $a<a_{0}$ have been proposed. 
This modified regime at $a<a_{0}$ will generically be characterised by equilibrium velocities which become independent of 
distance. Here we identify a critical test in this debate and we propose its application to samples of wide binary stars. 
Since for $1 M_{\odot}$ systems the acceleration drops below $a_{0}$ at scales of around 7000 AU, a statistical survey of 
wide binaries with relative velocities and separations reaching $10^{4}$ AU and beyond should prove useful to the above debate. 
We apply the proposed test to the best currently available data.
Results show a constant upper limit to the relative velocities in wide binaries which is 
independent of separation for over three orders of magnitude, in analogy with galactic flat rotation curves in the 
same $a<a_{0}$ acceleration regime. Our results are suggestive of a breakdown of Kepler's third law 
beyond $a \approx a_{0}$ scales, in accordance with generic predictions of modified gravity theories designed not to require 
any dark matter at galactic scales and beyond.
\end{abstract}

\section{Introduction}

Over the past years the dominant explanation for the large mass to light ratios inferred for galactic and meta-galactic 
systems, that these are embedded within massive dark matter halos, has begun to be challenged. Direct detection of the dark 
matter particles, in spite of decades of extensive and dedicated searches, remains lacking. Numerous alternative theories of 
gravity have appeared (e.g. TeVeS of \cite{3}, F(R) theories e.g. \cite{29}) mostly grounded on geometrical extensions 
to General Relativity, and leading to laws of gravity which in the low acceleration regime, mimic the MOdified 
Newtonian Dynamics (MOND) fitting formulas e.g. \cite{24}. Similarly, \cite{22} have explored MOND not as a modification to 
Newton's second law, but as a modified gravitational force law in the Newtonian regime, finding a good agreement with observed 
dynamics across galactic scales without requiring dark matter. Recently \cite{4} have constructed an $f(R)$ extension 
to general relativity which in the low velocity limit converges to the above approach.

A generic feature of all such modified gravity schemes is the appearance of an acceleration scale, $a_{0}$, above which 
classical gravity is recovered, and below which the dark matter mimicking regime appears. This results in a general 
prediction; all systems where $a>>a_{0}$ should appear as devoid of dark matter, and all 
systems where $a<<a_{0}$ should appear as dark matter dominated.  It is interesting 
that no $a>>a_{0}$ system has ever been detected where dark matter needs to be invoked. 
On the other hand, the latter condition furnishes a testable prediction in relation to the orbits of wide binaries. For test 
particles in orbit around a $1 M_{\odot}$ star, in circular orbits of radius $s$, the acceleration is expected to drop below 
$a_{0}\approx 1.2 \times 10 ^{-10} m/s^{2}$ for $s>$7000 AU$=3.4\times10^{-2} pc$. This provides a test for the dark matter/modified 
theories of gravity debate; the relative velocities of components of binary stars with large physical separations should 
deviate from Kepler's third law. Seen as an equivalent Newtonian force law, beyond $s \approx $7000 AU the gravitational force 
should gradually switch from the classical form of $F_{N}=GM/s^{2}$ to $F_{MG}=(G M a_{0})^{1/2}/s$, and the 
orbital velocity, $V^{2}/s =F$, should no longer decrease with separation, but settle at a constant value, dependent 
only on the total mass of the system through $V=(G M a_{0})^{1/4}$. Under modified gravity theories, 
binary stars with separations beyond 7000 AU should exhibit ``flat rotation curves'' and a 
``Tully-Fisher relation'', as galactic systems in the same acceleration regime do.

We apply this test to the binaries of two recent catalogues containing relative velocities and separations.
The two catalogues are entirely independent in their approaches. The first one, \cite{29}, comprises 280 systems 
from the {\it Hipparcos} satellite, while the second, \cite{11}, identifies 1,250 ones
from the Sloan Digital Sky Survey (SDSS). Details can be found in \cite{100}.

\section{Relative velocity distributions for wide binaries under Newtonian expectations}

The Newtonian prediction for the relative velocities of the two components of binaries having circular 
orbits, when plotted against the binary physical separation, $s$, is for a scaling of $\Delta V \propto s^{-1/2}$, essentially 
following Kepler's third law, provided the range of masses involved were narrow. 
In a relative proper motion sample however, only two components of the relative velocity appear, as velocity along the line of 
sight to the binary leads to no proper motion. Thus, orbital projection plays a part, with systems having orbital planes along 
the line of sight sometimes appearing as having no relative proper motions. A further effect comes from any degree of orbital 
ellipticity present; it is hence clear that the trend for $\Delta V \propto s^{-1/2}$ described above, will only provide an upper 
limit to the distribution of projected $\Delta V$ vs. $s$ expected in any real observed sample, even if only a narrow range of 
masses is included. One should expect a range of measured values of projected $\Delta V$ at a fixed observed projected $s$, all 
extending below the Newtonian limit, which for equal mass binaries in circular orbits gives $\Delta V_{N} =2 \left( \frac{G M}{s} \right)^{1/2}$.

Further, over time, the orbital parameters of binaries will evolve due to the effects of Galactic tidal forces and dynamical encounters 
with other stars in the field, specially in the case of wide 
binaries. To first order, one would expect little evolution for binaries tighter than the tidal limit of 1.7pc, and the 
eventual dissolution of wider systems. 
A very detailed study of all these points has recently appeared, \cite{16}. These authors numerically follow 
populations of 50,000 $1 M_{\odot}$ binaries in the Galactic environment, accounting for the evolution of the orbital parameters 
of each due to the cumulative effects of the Galactic tidal field at the Solar radius.  Also, the effects of close and long range 
encounters with other stars in the field are carefully included, to yield a present day distribution of separations and relative 
velocities for an extensive population of wide binaries, under Newtonian Gravity.

It is found that when many wide binaries cross their Jacobi radius, the two components remain fairly close by in both coordinate
and velocity space. Thus, in any real wide binary search 
a number of wide pairs with separations larger than their Jacobi radii will appear. Finally, \cite{16} obtain the 
RMS one-dimensional relative velocity difference, $\Delta V_{1D}$, projected along a line of sight, for the entire 
populations of binaries dynamically evolved over 10 Gyr to today, as plotted against 
the projected separation on the sky for each pair. The expected Keplerian fall of $\Delta V_{1D} \propto s^{-1/2}$ for 
separations below 1.7 pc is obtained, followed by a slight rise in $\Delta V_{1D}$ as wide systems cross the Jacobi radius 
threshold. $\Delta V_{1D}$ then settles at RMS values of $ \approx 0.1 km/s$. This represents the best currently available estimate of how 
relative velocities should scale with projected separations for binary stars (both bound and in the process of dissolving in 
the Galactic tides) under Newtonian gravity. 
We see that all we need is a large sample of relative proper motion and binary separation measurements to   
test the Newtonian prediction for the RMS values of the 1 dimensional relative velocities of \cite{16},
and to contrast the $\Delta V_{N} \propto s^{-1/2}$ and the $\Delta V_{MG} =cte.$ predictions for the upper envelope
of the $\Delta V$ vs. $s$ distributions.

\section{Observed wide binary Samples}

In the \cite{29} catalogue wide binaries are identified by assigning a probability above chance 
alignment to the systems by carefully comparing to the underlying background (and its variations) in a 5 dimensional 
parameter space of proper motions and spatial positions. We keep only binaries with a probability 
of non-chance alignment greater than $0.9$. The binary search criteria used by the authors requires that the proposed 
binary should have no near neighbours; the projected separation between the two components is thus always many times smaller 
than the typical interstellar separation. We use the reported distances to the primaries, where errors are smallest, 
to calculate projected $\Delta V$ and projected $s$ from the measured $\Delta \mu$ and $\Delta \theta$ values reported. 
Although the use of {\it Hipparcos} measurements guarantees the best available quality in 
the data, we have also removed all binaries for which the final signal to noise ratio in 
the relative velocities was lower than $0.3$.

We are left with a sample of 280 binaries, having distances to the Sun within 
$6<d<100$ in pc. The data show a perfectly flat upper envelope in a $\Delta V$ vs. projected $s$, \cite{100}.
The average signal to noise ratio for the data is 1.7, with an average error on $\Delta V$ of 0.83 $km/s$, 
which considering a $2 \sigma$ factor from the top of the distribution to the real underlying upper limit for the sample, 
results in 3 $km/s$ as our estimate of the actual physical upper limit in $\Delta V$.

The Sloan low mass wide pairs catalogue (SLoWPoKES) of \cite{11} contains a little over 1,200 wide 
binaries with relative proper motions for each pair, distances and angular separations. Also, extreme care was 
taken to include only physical binaries, with a full galactic population model used to exclude chance alignment 
stars using galactic coordinates and galactic velocities, resulting in an estimate of fewer than 2\% of false 
positives. This yields only isolated binaries with no neighbours within many times the internal binary separation. 
Again, we use the reported distances to the primaries to calculate 
projected $\Delta V$ and projected $s$ from the measured $\Delta \mu$, $\Delta \theta$ and $d$ values reported by \cite{11}, 
to obtain a sample of 417 binaries.

The upper envelope of the distribution of 
$\Delta V$ from this catalogue does not comply with Kepler's third law. As was the case with the {\it Hipparcos} 
sample, the upper envelope describes a flat line, as expected under modified gravity 
schemes. The average signal to noise in $\Delta V$ for the \cite{11} catalogue is
0.48, with an average error on $\Delta V$ of 12 $km/s$, which considering a $2 \sigma$ factor from the top 
of the complete distribution to the real underlying upper limit gives the same $3 km/s$ as obtained for the \cite{29} 
{\it Hipparcos} catalogue.


Figure (1) shows the RMS value of the one-dimensional relative velocity difference for 
both of the samples discussed. The error bars give the error propagation on $\Delta \mu$ and $d$. We construct $\Delta V_{1D}$ 
by considering only one coordinate of the two available from the relative motion on the 
plane of the sky. Thus, each binary can furnish two $\Delta V_{1D}$ measurements, which statistically should not introduce any 
bias. Indeed, using only $\Delta \mu_{l}$ or only $\Delta \mu_{b}$ or both for each binary, yields the same mean values for the 
points shown. The small solid error bars result from considering an enlarged sample where each binary contributes two $\Delta V_
{1D}$ measurements, while the larger dotted ones come from considering each binary only once, and do not change if we consider 
only $\Delta \mu_{l}$ or only $\Delta \mu_{b}$. The series of small $log(s)$ interval data are for the {\it Hipparcos} catalogue 
of \cite{29}, while the two broader crosses show results for the \cite{11} SDSS sample. 

The solid curve is the Newtonian prediction of the full Galactic evolutionary model of \cite{16} for 
binaries, both bound and in the process of dissolving. Note that the results of this simulation deviate from Kepler's law 
for $s$ larger than the Newtonian Jacobi radius of $1.7 pc$, whereas the discrepancy with
the observed samples also occurs at much smaller separations.  Even considering the large error bars, where 
each binary contributes only one $\Delta V_{1D}$ value, we see eight points lying beyond 1$\sigma$, making the probability of 
consistency between this prediction and the observations of less than $(0.272)^{8}$=$3\times 10^{-5}$.

We obtain a constant RMS value for $\Delta V_{1D}$ of 1 $km/s$, in qualitative agreement with expectations 
from modified gravity schemes. The vertical line marks $a=a_{0}$; we see the data departing from the 
Newtonian prediction outwards of this line, and not before.
The two independent catalogues, each using different sets of selection criteria, each 
perhaps subject to its own independent systematics, are consistent with the same result, a constant horizontal upper envelope for 
the distribution of relative velocities on the plane of the sky at an intrinsic value of 3 $km/s \pm$1 $km/s$, extending over 3 
orders of magnitude in $s$, with a constant RMS $\Delta V_{1D}$ value consistent with 1 $km/s \pm$0.5 $km/s$. This supports the 
interpretation of the effect detected as the generic prediction of modified gravity theories.

\begin{figure}[h]
\includegraphics[width=20pc]{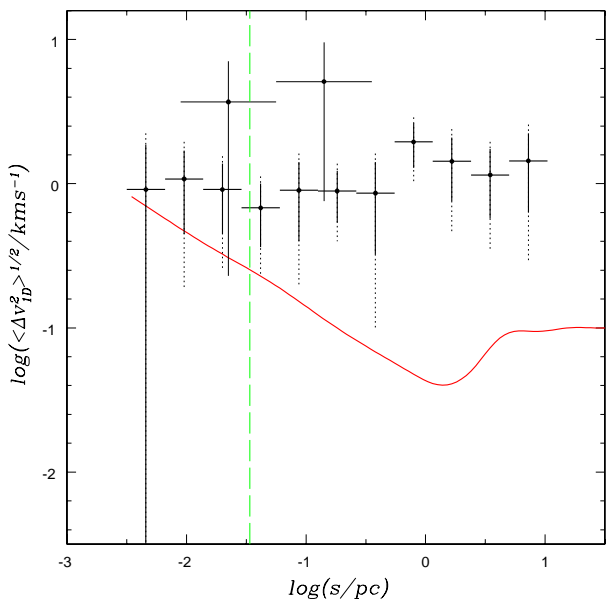}\hspace{2pc}%
\begin{minipage}[b]{14pc}\caption{\label{label}
The solid curve gives the RMS values for one dimensional projected relative velocities as a function of
projected separations, for the dynamical modelling of large populations of wide binaries evolving in
the Galactic environment, taken from \cite{16}. The same quantity for the data from the catalogues analysed is
given by the points with error bars; those with narrow $log(s)$ intervals being from the {\it Hipparcos} sample of \cite{29},
and those two with wide $log(s)$ intervals coming from the SDSS sample of \cite{11}}
\end{minipage}
\end{figure}

\section{Conclusions}

We identify a critical test in the classical gravity/modified gravity debate, using the relative velocities of wide 
binaries with separations in excess of 7000 AU, as these occupy the $a<a_{0}$ regime.
We present a first application of this test using the best currently available data.
Results show constant relative RMS velocities for the binary stars in question. This is quantitatively inconsistent 
with detailed predictions of Newtonian dynamical models for binaries evolving in the local galactic environment.
Our results are qualitatively in accordance with generic modified gravity models which explain galactic dynamics in the 
absence of dark matter.


\end{document}